\documentstyle[12pt,plenum]{book}
\makeatletter

\def\@plus{plus}
\makeatother
\input psfig
\newcommand{\site}[1]{\refnote{\cite{#1}}}
\newcommand{\beq}{\begin{equation}}
\newcommand{\eeq}{\end{equation}}
\newcommand{\bea}{\begin{eqnarray}}
\newcommand{\eea}{\end{eqnarray}}
\newcommand{\un}[1]{{\it #1}}
\newcommand{\half}{{\scriptstyle{{1\over 2}}}}

\newcommand{\real}{\relax{\rm I\kern-.18em R}}
\newcommand{\zahlen}{{\rm Z \!\! Z}}

\newcommand{\quat}{{\rm I \! H}}

\newcommand{\id}{\mbox{$id$}}

\newcommand{\Tr}{\mbox{\,Tr\,}}
\newcommand{\ad}{{\rm ad}}

\newcommand{\norm}[1]{\left\| #1 \right\|}
\newcommand{\sgbar}{\bar{\sg}}
\newcommand{\tr}{\mbox{\,tr\,}}
\newcommand{\al}{\alpha}
\newcommand{\Gm}{\Gamma}

\newcommand{\dl}{\delta}

\newcommand{\veps}{\varepsilon}

\newcommand{\lm}{\lambda}
\newcommand{\Lm}{\Lambda}
\newcommand{\sg}{\sigma}

\newcommand{\Om}{\Omega}
\newcommand{\Ss}[1]{\mbox{$\cal #1$}}
\newcommand{\pr}{\partial}

\newcommand{\Lkw}{\vec{L}_1^2}
\newcommand{\ip}[1]{\left\langle #1 \right \rangle }
\newcommand{\Order}[1]{\Ss{O}\left(#1\right)}

\begin{document}
\chapter{
{}\vskip-3cm\hfill{\tenrm INLO-PUB-10/97}\vskip2cm 
GRIBOV AMBIGUITIES AND THE FUNDAMENTAL DOMAIN}

\author{Pierre van Baal\refnote{1,2}}

\affiliation{\affnote{1}Isaac Newton Institute for Mathematical Sciences,\\
20 Clarkson Road, Cambridge CB3 0EH, UK\\
\affnote{2}Instituut-Lorentz for Theoretical 
Physics\footnote{Permanent address. 
Lectures deliverd at the NATO ASI ``Confinement, Duality and Non-perturbative 
Aspects of QCD'', Newton Institute, Cambridge, UK, 23 June - 4 July, 1997.
}, University of Leiden,\\ 
P.O.Box 9506, NL-2300 RA Leiden, The Netherlands
}

\vskip1.5cm
\hskip1.6cm{\em DEDICATED TO THE MEMORY OF VLADIMIR GRIBOV}

\section{INTRODUCTION}
Non-perturbative aspects are believed to play a crucial role in understanding 
the formation of a mass gap in the spectrum of excitations in a non-Abelian 
gauge theory, but despite much progress a simple explanation is still lacking. 
Here this problem will be addressed in a finite volume, where its size can be 
used as a control parameter, which is absent in infinite volumes. The starting 
point, as in solving a quantum mechanical problem, is finding a proper 
description of the classical potential and its minima.

Due to the non-Abelian nature of the theory not only the classical potential, 
but also the kinetic term in the Lagrangian or Hamiltonian, is more complicated.
The latter is a manifestation of the non-trivial Riemannian geometry of the 
physical configuration space\site{Bavi}, formed by the set of gauge orbits 
$\cal{A}/\cal{G}$ ($\cal{A}$ is the collection of connections, $\cal{G}$ the 
group of local gauge transformations). Most frequently, coordinates of this 
orbit space are chosen by picking a representative gauge field on the orbit 
in a smooth and preferably unique way. It is by now well known that linear 
gauge conditions like the Landau or Coulomb gauge suffer from Gribov 
ambiguities\site{Grib}. The reason behind this is that topological obstructions
prevent one from introducing affine coordinates\site{Sing} in a global way. In 
principle therefore, one can introduce different coordinate patches with 
transition functions to circumvent this problem\site{Nahm}. One way to make 
this specific is to base the coordinate patches on the choice of a background 
gauge condition\site{Kovb,Vba3}. One could envisage to associate to each 
coordinate patch ghost fields and extend BRST symmetry to include fields with 
non-trivial ``Grassmannian sections'', although such a formulation is still in 
its infancy. 

We will pursue, however, the issue of finding a fundamental domain for
non-Abelian gauge theories\site{Sefr} and its consequence for the glueball
spectrum in intermediate volumes. The finite volume context allows us to make
reliable statements on the non-perturbative contributions, because asymptotic
freedom guarantees that at small volumes the effective coupling constant is
small, such that high-momentum states can be treated perturbatively. Only
the lowest (typically zero or near-zero momentum) states will be affected by
non-perturbative corrections. We emphasize that it is essential that gauge
invariance is implemented properly at all stages. We will describe the results
mainly in the context of a Hamiltonian picture\site{Chle} with wave functionals
on configuration space. Although rather cumbersome from a perturbative point
of view, where the covariant path integral approach of Feynman is vastly
superior, it provides more intuition on how to deal with non-perturbative
contributions to observables that do not vanish in perturbation theory.
An essential feature of the non-perturbative behaviour is that the wave
functional spreads out in configuration space to become sensitive to its
non-trivial geometry. If wave functionals are localised within regions
much smaller than the inverse curvature of the field space, the curvature
has no effect on the wave functionals.  At the other extreme, if the
configuration space has non-contractible circles, the wave functionals
are drastically affected by the geometry, or topology, when their support
extends over the entire circle. Instantons are of course the most important
examples of this. Not only the vacuum energy is affected by these
instantons, but also the low-lying glueball states and this is what we
are after to describe accurately, albeit in sufficiently small volumes.
The geometry of the finite volume, to be considered here, is the one
of a three-torus\site{Lue1,Kovb} and a three-sphere\site{Cutk,Vbda}.
These lecture notes are an updated and extended version of ref~\site{Tren}.

\section{COMPLETE GAUGE FIXING}

An (almost) unique representative of the gauge orbit is found by
minimising the $L^2$ norm of the vector potential along the gauge
orbit\site{Sefr,Del1}
\beq
F_A(g)~\equiv\norm{^g A}^2~=~ -\int_M d^3x~
\tr \left( \left( g^{-1} A_i g + g^{-1} \pr_i g \right)^2\right),
\label{gAnorm}
\end{equation}
where the vector potential is taken anti-hermitian. 
Expanding around the minimum of eq.~(\ref{gAnorm}), writing $g(x)=\exp(X(x))$,
one easily finds:
\bea
\norm{^g A}^2 &=& \norm{A}^2+2\int_M \tr(X
\partial_i A_i)+\int_M \tr (X^\dagger FP (A) X) \nonumber \\
&&+\frac{1}{3}\int_M\tr\left(X\left[[A_i,X],\partial_i X\right]\right)
+\frac{1}{12}\int_M\tr\left([D_iX,X][\partial_i X,X]\right)+\Order{X^5}.
\label{Xexpansie}
\eea
Where $FP(A)$ is the Faddeev-Popov operator $(\ad(A)X~\equiv~[A,X])$
\beq
FP (A)~=~-\partial_i D_i (A)~=~-\partial^2_i -\partial_i\ad(A_i).
\label{FPdef}
\eeq

At any local minimum the vector potential is therefore transverse,
$\partial_i A_i~=~0$, and $FP(A)$ is a positive operator. The
set of all these vector potentials is by definition the Gribov region
$\Omega$. Using the fact that $FP(A)$ is linear in $A$, $\Omega$ is
seen to be a convex subspace of the set of transverse connections $\Gamma$.
Its boundary $\partial \Omega$ is called the Gribov horizon. At the Gribov
horizon, the \un{lowest} eigenvalue of the Faddeev-Popov operator
vanishes, and points on $\partial\Omega$ are hence associated with coordinate
singularities. Any point on $\partial\Omega$ can be seen to have a finite
distance to the origin of field space and in some cases even
uniform bounds can be derived\site{Dezw,Zwan}.

The Gribov region is the set of \un{local} minima of the
norm functional (3) and needs to be further restricted to the
\un{absolute} minima to form a fundamental domain, which will be denoted
by $\Lambda$. The fundamental domain is clearly contained within the
Gribov region. To show that also $\Lambda$ is convex we note that
\bea
  \norm{^gA}^2 - \norm{A}^2 &=&
     \int \tr \left( A_i^2 \right)
     - \int \tr \left( \left( g^{-1} A_i g + g^{-1} \pr_i g \right)^2\right)
   \nonumber \\
  &=& \int \tr \left( g^\dagger FP_f(A)~g \right)
       \equiv \ip{g,FP_f(A)~g},
  \label{FPhalfdef}
\eea
where $FP_f(A)$  is the Faddeev-Popov operator generalised to the fundamental 
representation. At the critical points $A \in \Gm$ of the norm functional
(recall $\Gm = \{ A \in \Ss{A} | \pr_i A_i = 0 \}$) $FP_f(A)$, like $FP(A)$, 
is a hermitian operator. We can define $\Lm$ in terms of the absolute minima 
over $g \in \Ss{G}$ of $\ip{g, FP_f(A)~g}$
\beq
  \Lm =
  \{ A \in \Gm | \min_{g \in \Ss{G}} \ip{g, FP_f(A)~g}=0 \}.
  \label{Lmdef}
\eeq
Using that $FP_f(A)$ is linear in $A$ and assuming that $A^{(1)}$ and
$A^{(2)}$ are in $\Lm$ and therefore satisfy the equation
$\min_{g \in \Ss{G}} \ip{g, FP_f(A)~g}=0$, we find that 
$A=sA^{(1)}+(1-s)A^{(2)}$ satisfies the same identity for all $s\in[0,1]$ 
(such that both $s$ and $(1-s)$ are positive). The line connecting two 
points in $\Lm$, therefore lies within $\Lm$.

If we would not specify anything further, as a convex space is contractible,
the fundamental region could never reproduce the non-trivial topology of the
configuration space. This means that $\Lm$ should have a boundary\site{Vba1}. 
Indeed, as $\Lambda$ is contained in $\Omega$, this means $\Lm$ is also 
bounded in each direction. Clearly $A=0$ is in the interior of $\Lm$, which
allows us to consider a ray extending out from the origin into a given 
direction, where it will have to cross the boundary of $\Lm$ and $\Om$.
For any point along this ray in $\Lm$, the norm functional is at its 
absolute minimum as a function of the gauge orbit. However, for points in
$\Om$ that are not also in $\Lm$, the norm functional is necessarily at
a relative minimum. The absolute minimum for this orbit is an element
of $\Lm$, but in general not along the ray. Continuity therefore tells
us that at some point along the ray, this absolute minimum has to pass
the local minimum. At the point they are exactly degenerate, there are
two gauge equivalent vector potentials with the same norm, both at the 
absolute minimum. As in the interior the norm functional has a unique
minimum, again by continuity, these two degenerate configurations have
to both lie on the boundary of $\Lm$. This is the generic situation.

It is important to note that $A=0$ is a so-called reducible 
connection\site{Dona} which has a non-trivial stabiliser, i.e. a subgroup of 
the gauge group that leaves the connection invariant, in this case the set of 
constant gauge transformations. These reducible connections give rise to 
so-called orbifold singularities, that manifest themselves through curvature 
singularities. We know perfectly well how to deal with it by not fixing
the constant gauge transformations. Indeed the norm functional is degenerate
along the constant gauge transformations and the Coulomb gauge does not fix
this gauge degree of freedom. We simply demand that the wave functional is in
the singlet representation under the constant gauge transformations.
The degeneracy of the norm functional along the constant gauge transformations 
gives rise to a trivial part in the kernel of the Faddeev-Popov operator (also 
directly seen by inspection, since it vanishes when acting on constant 
Lie-algebra elements that generate the constant gauge transformations). 
This trivial kernel is to be projected out, something that is possible in a 
finite volume. In this formalism there is a remnant gauge invariance $G$, 
which requires no gauge fixing since its volume is finite, but still needs 
to be divided out to get the proper identification
\beq
\Lm/G~=~ \cal{A}/\cal{G}.
\eeq
Here $\Lm$ is assumed to include the non-trivial boundary identifications.
It is these boundary identifications that restore the non-trivial topology of 
$\cal{A}/\cal{G}$. 

If the degeneracy at the boundary is continuous along non-trivial directions
one necessarily has at least one non-trivial zero eigenvalue for $FP(A)$ and 
the Gribov horizon will touch the boundary of the fundamental domain at these 
so-called singular boundary points. We sketch the general situation in figure 
1. By singular we mean here a coordinate singularity. In principle, by choosing
a different gauge fixing in the neighbourhood of these points one could resolve 
the singularity. If singular boundary points would not exist, all that would 
have been required is to complement the Hamiltonian in the Coulomb gauge with 
the appropriate boundary conditions in field space. Since the boundary 
identifications are by gauge transformations the boundary condition on the 
wave functionals is simply that they are identical under the boundary 
identifications, possibly up to a phase in case the gauge transformation is 
homotopically non-trivial, as will be discussed further on in more detail. 

\begin{figure*}{\tt}
\vspace{11.7cm}
\includegraphics{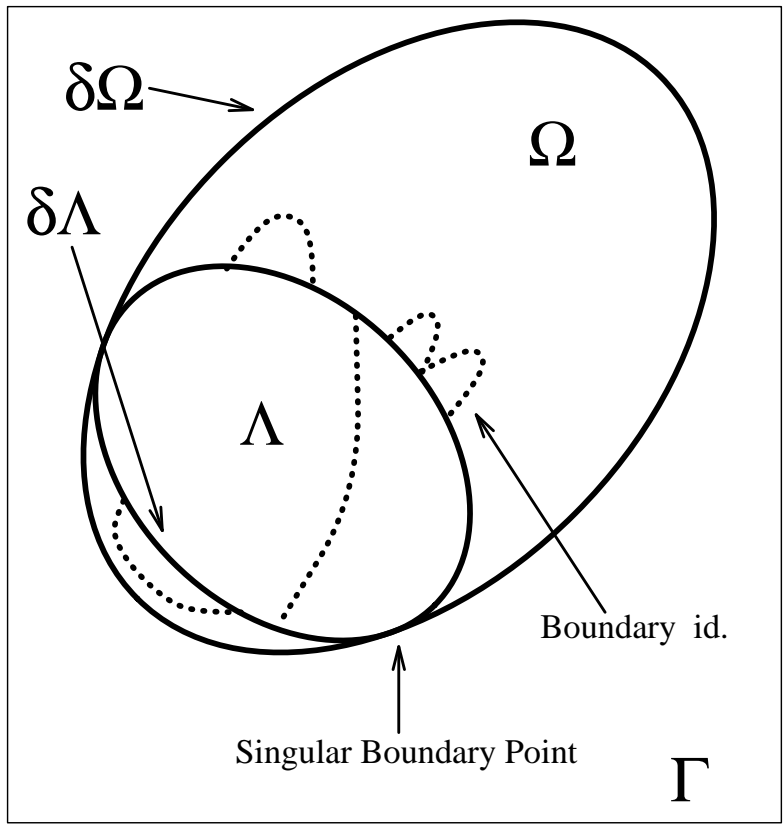}
\caption{Sketch of the Gribov and fundamental regions. The dotted lines
indicate the boundary identifications.}
\label{fig:fig1}
\end{figure*}

Unfortunately, the existence of non-contractible spheres\site{Sing} in the
configuration space allows one to argue that singular boundary points are to 
be expected\site{Vba1}. Consider the intersection of a $d$-dimensional 
non-contractible sphere with $\Lm$. The part in the interior of $\Lm$ is 
contractible and it can only become non-contractible through the boundary 
identifications. The simplest way this can occur\site{Vba1} is if for the 
($d$-1)-dimensional intersection with the boundary, all points are to be 
identified. It would imply degeneracy of the the norm functional on a 
($d$-1)-dimensional subspace, leading to at least $d$-1 zero-modes for the 
Faddeev-Popov operator. The intersection with the boundary of $\Lm$ can, 
however, exist of more than one connected component. In the case of two such 
components one can make a non-contractible sphere by identifying points of 
the ($d$-1)-dimensional boundary intersection of the first connected component 
with that of the second and there is no necessity for a continuous degeneracy. 
Conversely, not all singular boundary points, even those associated with 
continuous degeneracies, need to be associated with non-contractible spheres. 

When a singular boundary point is not associated to a continuous degeneracy, 
the norm functional undergoes a bifurcation moving from inside to outside the 
fundamental (and Gribov) region. The absolute minimum turns into a 
saddle point and two local minima appear, as indicated in figure 2. These 
are necessarily gauge copies of each other. The gauge transformation is 
homotopically trivial as it reduces to the identity at the bifurcation point, 
evolving continuously from there on. For reducible connections, that have a 
non-trivial stabiliser, this argument may be false\site{Vba2}, but examples of 
bifurcations at irreducible connections were explicitly found for $S^3$, see 
ref.\site{Vbvd} (app. A). We will come back to this.

\begin{figure*}{\tt}
\vspace{13.2cm}
\includegraphics{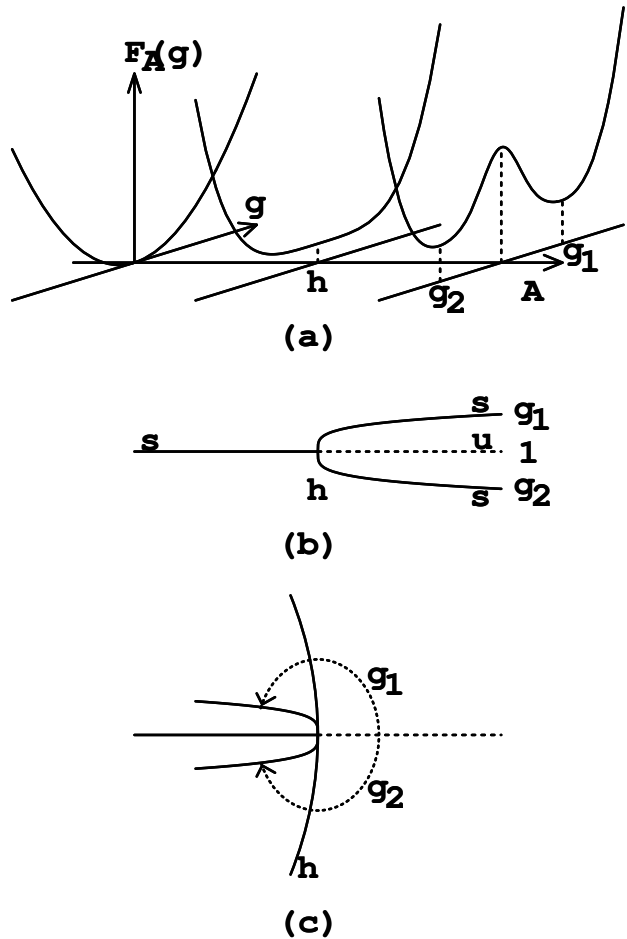}
\caption{Sketch of a singular boundary point due to a bifurcation of
the norm functional. It can be used to show that there are homotopically
trivial gauge copies inside the Gribov horizon ($h$).}
\label{fig:fig2}
\end{figure*}

Also Gribov's original arguments for the existence of gauge copies\site{Grib}
(showing that points just outside the horizon are gauge copies of points
just inside) can be easily understood from the perspective of bifurcations in
the norm functional. It describes the generic case where the zero-mode of
the Faddeev-Popov operator arises because of the coalescence of a local
minimum with a saddle point with only one unstable direction. At the Gribov
horizon the norm functional locally behaves in that case as $X^3$, with $X$
the relevant zero eigenfunction of the Faddeev-Popov operator. The situation
sketched in figure 2 corresponds to the case where the leading behaviour is
like $X^4$. See ref.\site{Vba1} for more details and a discussion of the Morse 
theory aspects that simplify the bifurcation analysis. 
As the Gribov region is associated with the local minima, and since the space 
of gauge transformations resembles that of a spin model, the analogy with spin 
glasses makes it unreasonable to expect that the Gribov region is free of 
further gauge copies. This will be illustrated by explicit examples. 
Unfortunately restrictions to a subset of the transverse gauge fields is 
a rather non-local procedure. This cannot be avoided since it reflects the 
non-trivial topology of field space.

\section{GAUGE FIELDS ON THE THREE-TORUS}

Homotopical non-trivial gauge transformations are in one to one
correspondence with non-contractible loops in configuration space, which
give rise to conserved quantum numbers. The quantum numbers are like the
Bloch momenta in a periodic potential and label representations of the
homotopy group of gauge transformations. On the fundamental domain
the non-contractible loops arise through identifications of boundary points
(as will be demonstrated quite explicitly for the torus in the
zero-momentum sector). Although slightly more hidden, the fundamental
domain will therefore contain all the information relevant
for the topological quantum numbers. Sufficiently accurate knowledge of the
boundary identifications will allow for an efficient and natural
projection on the various superselection sectors (i.e. by choosing
the appropriate ``Bloch wave functionals''). All these features
were at the heart of the finite volume analysis on the torus\site{Kovb} and
we see that they can in principle be naturally extended to the
full theory, thereby including the desired $\theta$ dependence. In the
next section this will be discussed in the context of the three-sphere.
In ref.\site{Vba3} we proposed formulating the Hamiltonian theory on
coordinate patches, with homotopically non-trivial gauge transformations as
transition functions. Working with boundary conditions on the boundary
of the fundamental domain is easily seen to be equivalent and conceptually
much simpler to formulate. If there would be no singular boundary points
this would have provided a Hamiltonian formulation where all topologically
non-trivial information can be encoded in the boundary conditions. Still,
for the low-lying states in a finite volume, both on the three-torus and
the three-sphere, singular boundary points will not play an important role
in intermediate volumes.

Probably the most simple example to illustrate the relevance of the
fundamental domain is provided by gauge fields on the torus in the
abelian zero-momentum sector. For definiteness let us take $G=SU(2)$
and $A_i=i{C_i\over 2L}\tau_3$ ($L$ is the size of the torus). These
modes are dynamically motivated as they form the set of gauge
fields on which the classical potential vanishes. It is called the
vacuum valley (sometimes also referred to as toron valley) and one can
attempt to perform a Born-Oppenheimer-like approximation for deriving
an effective Hamiltonian in terms of these ``slow'' degrees of freedom.
To find the Gribov horizon, one easily verifies that the part of the spectrum
for $FP(A)$ that depends on $\vec C$, is given by $\lambda^{gh}_{\vec n}(\vec
C)=2\pi\vec n\cdot(2\pi\vec n\pm\vec C)$, with $\vec n\neq\vec 0$ an integer
vector. The lowest eigenvalue therefore vanishes if $C_k=\pm2\pi$. The Gribov
region is therefore a cube with sides of length $4\pi$, centred at the origin,
specified by $|C_k|\leq2\pi$ for all $k$, see figure 3.

The gauge transformation $g_{(k)}=\exp(\pi i x_k\tau_3/L)$ maps $C_k$ to
$C_k+2\pi$, leaving the other components of $\vec C$ untouched. As $g_{(k)}$
is anti-periodic it is homotopically non-trivial (they are 't Hooft's twisted
gauge transformations\site{Tho1}). We thus see explicitly that gauge copies
occur inside $\Om$, but furthermore the naive vacuum $A=0$ has (many) gauge
copies under these shifts of $2\pi$ that lie on the Gribov horizon. It can
actually be shown for the Coulomb gauge that for any three-manifold, any
Gribov copy by a homotopically non-trivial gauge transformation of $A=0$ will
have vanishing Faddeev-Popov determinant\site{Vba1}. Taking the symmetry
under homotopically non-trivial gauge transformations properly into account
is crucial for describing the non-perturbative dynamics and one sees that
the singularity of the Hamiltonian at Gribov copies of $A=0$, where the
wave functionals are in a sense maximal, could form a severe obstacle in
obtaining reliable results.

To find the boundary of the fundamental domain we note that the gauge copies
$\vec C=(\pi,C_2,C_3)$ and $\vec C=(-\pi,C_2,C_3)$ have equal norm. The
boundary of the fundamental domain, restricted to the vacuum valley formed by
the abelian zero-momentum gauge fields, therefore occurs where $|C_k|=\pi$,
well inside the Gribov region, see figure 3. The boundary identifications are
by the homotopically non-trivial gauge transformations $g_{(k)}$. The
fundamental domain, described by $|C_k|\leq \pi$, with all boundary points
regular, has the topology of a torus. To be more precise, as the remnant of
the constant gauge transformations (the Weyl group) changes $\vec C$ to
$-\vec C$, the fundamental domain $\Lm/G$ restricted to the abelian constant
modes is the orbifold $T^3/Z_2$. Generalisations to arbitrary gauge groups
were considered in ref.\site{Vba3}. (The fundamental domain turns out to
coincide with the unit cell or ``minimal'' coordinate patch defined in
ref.\site{Vba3}).

\begin{figure}{\tt}
\vspace{7.5cm}
\includegraphics{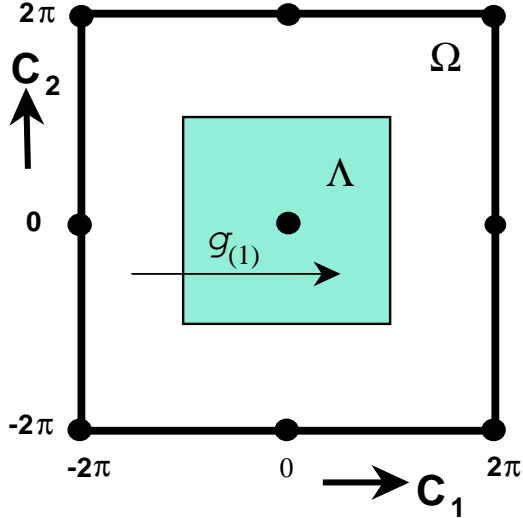}
\caption{A two dimensional slice of the vacuum valley along the $(C_1,C_2)$
plane. The fat square give the Gribov horizon, the grey square is the
fundamental domain. The dots at the Gribov horizon are Gribov copies
of the origin.}
\label{fig:fig3}
\end{figure}

Formulating the Hamiltonian on $\Lm$, with the boundary identifications
implied by the gauge transformations $g_{(k)}$, avoids the singularities
at the Gribov copies of $A=0$. ``Bloch momenta'' associated to the $2\pi$
shift, implemented by the non-trivial homotopy of $g_{(k)}$, label
`t Hooft's electric flux quantum numbers\site{Tho1} $\Psi(C_k=-\pi)=\exp(
\pi ie_k)\Psi(C_k=\pi)$. Note that the phase factor is not arbitrary, but
$\pm 1$. This is because $g^2_{(k)}$ is homotopically trivial. In other
words, the homotopy group of these anti-periodic gauge transformations
is $Z_2^3$. Considering a slice of $\Lambda$ can obscure some of the
topological features. A loop that winds around the slice twice is
contractible in $\Lambda$ as soon as it is allowed to leave the slice.
Indeed including the lowest modes transverse to this slice will make the
$Z_2$ nature of the relevant homotopy group evident\site{Kovb}. It should
be mentioned that for the torus in the presence of fields in the fundamental
representation (quarks), only periodic gauge transformations are allowed.
In that case it is easily seen that the intersection of the fundamental
domain with the constant abelian gauge fields is given by the domain
$|C_k|\leq 2\pi$, whose boundary coincides with the Gribov horizon. It is
interesting to note that points on $\partial\Om$ form an explicit example of
a continuous degeneracy due to a non-contractible sphere\site{Zwan}.

In weak coupling L\"uscher\site{Lue1} showed unambiguously that the
wave functionals are localised around $A=0$, that they are normalisable and
that the spectrum is discrete. In this limit the spectrum is insensitive to
the boundary identifications (giving rise to a degeneracy in the topological
quantum numbers). This is manifested by a vanishing electric flux energy,
defined by the difference in energy of a state with $|\vec e|=1$ and the
vacuum state with $\vec e=\vec 0$. Although there is no classical potential
barrier to achieve this suppression, it comes about by a quantum induced
barrier, in strength down by two powers of the coupling constant. This gives
a suppression\site{Vbk1} with a factor $\exp(-S/g)$ instead of the usual factor
of $\exp(-8\pi^2/g^2)$ for instantons\site{Tho2}. Here $S=12.4637$ is the
action computed from the effective potential. At stronger coupling the wave
functional spreads out over the vacuum valley and the boundary conditions
drastically change the spectrum\site{Kovb}. At this point the energy of
electric flux suddenly switches on.

Integrating out the non-zero momentum degrees of freedom, for which Bloch 
degenerate perturbation theory provides a rigorous framework\site{Bloc,Lue1}, 
one finds an effective Hamiltonian. Near $A=0$, due to the quartic nature of 
the potential energy $(F^a_{ij})^2$ for the zero-momentum modes (the 
derivatives vanish and the field strength is quadratic in the field), there 
is no separation in time scales between the abelian and non-abelian modes. 
Away from $A=0$ one could further reduce the dynamics to one along the vacuum 
valley, but near the origin this would be a singular decomposition (the 
adiabatic approximation breaks down). However, as long as the coupling 
constant is not too large, the wave functional can be reduced to a wave 
function on the vacuum valley near $\partial\Lm$ where the boundary 
conditions can be implemented. These boundary conditions are
formulated in a manner that preserves the invariance under constant
gauge transformation and the effective Hamiltonian is solved by Rayleigh-Ritz
(providing \un{also} lower bounds from the second moment of the Hamiltonian).
The influence of the boundary conditions on the low-lying glueball states
is felt as soon as the volume is bigger than an inverse scalar glueball
mass. We summarise below the ingredients that enter the calculations.

The effective Hamiltonian is expressed in terms of the coordinates $c_i^a$,
where $i=\{1,2,3\}$ is the spatial index ($c_0=0$) and $a=\{1,2,3\}$ is
the SU(2)-colour index. These coordinates are related to the zero-momentum
gauge fields through $A_i^a(x)=c_i^a/L$. We note that the field strength
is given by $ F_{ij}^a=-\varepsilon_{abd}c_i^bc_j^d/L^2$ and we introduce
the gauge-invariant ``radial'' coordinate $r_i=\sqrt{\sum_a c_i^ac_i^a}$.
The latter will play a crucial role in specifying the boundary
conditions. For dimensional reasons the effective Hamiltonian is proportional
to $1/L$. It will furthermore depend on $L$ through the renormalised coupling
constant ($g(L)$) at the scale $\mu=1/L$. To one-loop order one has (for small
$L$) $g(L)^2=12\pi^2/[-11\ln(\Lambda_{MS}L)]$. One expresses the masses and
the size of the finite volume in dimensionless quantities, like mass-ratios
and the parameter $z=mL$. In this way, the explicit dependence of $g$ on $L$
is irrelevant. This is also the preferred way of comparing results obtained
within different regularisation schemes (i.e. dimensional and lattice
regularisation). The effective Hamiltonian is now given by
\bea
L\cdot H_{eff}(c)&=&{g^2\over 2(1+\alpha_1 g^2)}\sum_{i,a}
{\partial^2\over \partial
{c_i^a}^2}+{1\over 4}({1\over g^2}+\alpha_2)\sum_{ij,a}{F_{ij}^a}^2\nonumber\\
&&+\gamma_1\sum_i r_i^2+\gamma_2\sum_i r_i^4+\gamma_3\sum_{i>j} r_i^2r_j^2+
\gamma_4\sum_i r_i^6+\gamma_5\sum_{i\neq j} r_i^2 r_j^4
+\gamma_6\prod_i r_i^2\nonumber\\
&&+\alpha_3\sum_{ijk,a}r_i^2{F_{jk}^a}^2+\alpha_4\sum_{ij,a}r_i^2{F_{ij}^a}^2
+\alpha_5{\det}^2c.
\eea

We have organised the terms according to the importance of their contributions,
ignoring terms quartic in the momenta. The first line gives (when ignoring
$\alpha_{1,2}$) the lowest order effective Hamiltonian, whose energy
eigenvalues are ${\cal O}(g^{2/3})$, as can be seen by rescaling $c$ with
$g^{2/3}$. Thus, in a perturbative expansion $c={\cal O}(g^{2/3})$. The
second line includes the vacuum-valley effective potential (i.e. the part
that does not vanish on the set of abelian configurations). These two lines
are sufficient to obtain the mass-ratios to an accuracy of better than
5\%. The third line gives terms of ${\cal O}(g^4)$ in the effective potential,
that vanish along the vacuum-valley. The coefficients (to two-loop order
for $\gamma_i$) are
\bea
\gamma_1=-3.0104661\cdot10^{-1}-(g/2\pi)^23.0104661\cdot10^{-1}&,&
\alpha_1=+2.1810429\cdot10^{-2},
\nonumber\\
\gamma_2=-1.4488847\cdot10^{-3}-(g/2\pi)^29.9096768\cdot10^{-3}&,&
\alpha_2=+7.5714590\cdot10^{-3},
\nonumber\\
\gamma_3=+1.2790086\cdot10^{-2}+(g/2\pi)^23.6765224\cdot10^{-2}&,&
\alpha_3=+1.1130266\cdot10^{-4},
\nonumber\\
\gamma_4=+4.9676959\cdot10^{-5}+(g/2\pi)^25.2925358\cdot10^{-5}&,&
\alpha_4=-2.1475176\cdot10^{-4},
\nonumber\\
\gamma_5=-5.5172502\cdot10^{-5}+(g/2\pi)^21.8496841\cdot10^{-4}&,&
\alpha_5=-1.2775652\cdot10^{-3},
\nonumber\\
\gamma_6=-1.2423581\cdot10^{-3}-(g/2\pi)^25.7110724\cdot10^{-3}&.&
\eea

The choice of boundary conditions, associated to each of the irreducible
representations of the cubic group $O(3,\zahlen)$ and to the states that
carry electric flux\site{Tho1}, is best described by observing that the
cubic group is the semidirect product of the group of coordinate permutations
$S_3$ and the group of coordinate reflections $Z_2^3$. We denote the parity
under the coordinate reflection $c_i^a\rightarrow -c_i^a$ by $p_i=\pm 1$
($i$ fixed). The electric flux quantum number for the same direction will
be denoted by $q_i=\pm 1$.  This is related to the more usual additive
(mod 2) quantum number $e_i$ by $q_j=\exp(i\pi e_j)$. Note that for SU(2)
electric flux is invariant under coordinate reflections. If not all of the
electric fluxes are identical, the cubic group is broken to $S_2\times
Z_2^3$, where $S_2(\sim Z_2)$ corresponds to interchanging the two
directions with identical electric flux (unequal to the other electric flux).
If all the electric fluxes are equal, the wave functions are irreducible
representations of the cubic group. These are the four singlets $A_{1(2)}^\pm$,
which are completely (anti-)symmetric with respect to $S_3$ and have
each of the parities $p_i=\pm 1$. Then there are two doublets $E^\pm$,
also with each of the parities $p_i=\pm 1$ and finally one has four triplets
$T_{1(2)}^\pm$. Each of these triplet states can be decomposed into
eigenstates of the coordinate reflections.  Explicitly, for $T_{1(2)}^\pm$
we have one state that is (anti-)symmetric under interchanging the two-
and three-directions, with $p_2=p_3=-p_1=\mp 1$. The other two states are
obtained through cyclic permutation of the coordinates. Thus, any
eigenfunction of the effective Hamiltonian with specific electric flux
quantum numbers $q_i$ can be chosen to be an eigenstate of the parity
operators $p_i$. The boundary conditions of these eigenfunctions
$\Psi_{\vec q,\vec p}(c)$ are simply given by
\bea
\Psi_{\vec q,\vec p}(c)|_{_{r_i=\pi}}=0\quad,&\quad{\it if}\quad
p_iq_i=-1 \nonumber\\
{\partial\over\partial r_i}(r_i\Psi_{\vec q,\vec p}(c))|_{_{r_i=\pi}}=0\quad,
&\quad{\it if}\quad p_iq_i=+1
\eea
and one easily shows that with these boundary conditions the Hamiltonian is
hermitian with respect to the innerproduct $<\Psi,\Psi^\prime>=\int_{r_i
\leq\pi}d^9c\Psi^*(c)\Psi^\prime(c)$. For negative parity states ($\prod_i p_i
=-1$) this description is, however, not accurate\site{Voh1} as parity
restricted to the vacuum valley is equivalent to a Weyl reflection
(a remnant of the invariance under constant gauge transformations).

\begin{figure}{\tt}
\vspace{9cm}
\includegraphics{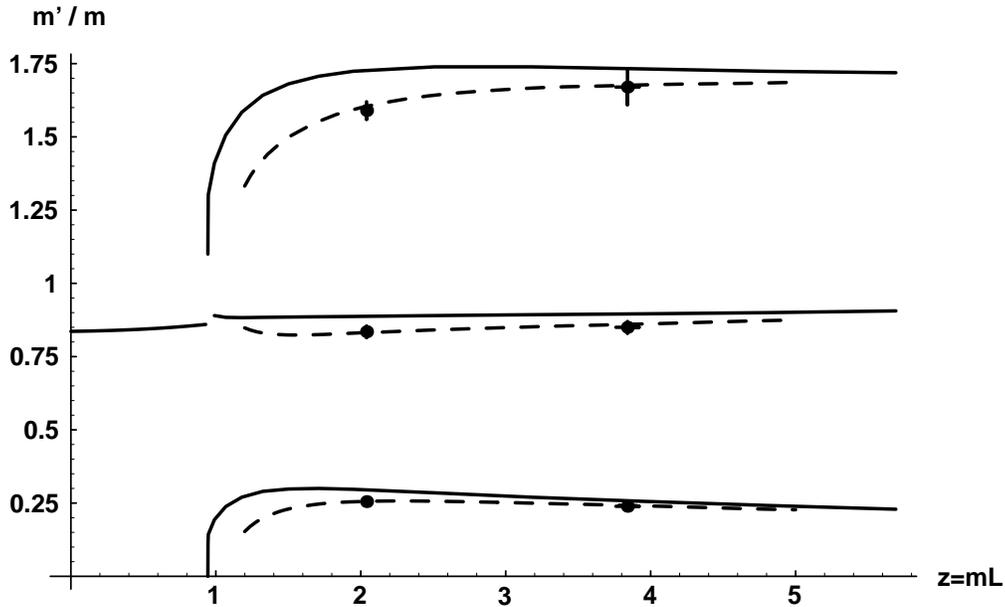}
\caption{Mass ratios $m^\prime/m$ as a function of $L$ in units of the inverse
scalar glueball mass $m^{-1}$, $z=mL$. The analytic results are given
by the full (continuum) and dashed ($4^3$ lattice) curves. Represented are
the the square root of the energy of electric flux per unit length (bottom),
the $E$ tensor mass (middle) and the $T_2$ tensor mass (top). The latter
two are (almost) degenerate below $z=0.95$, which is also where the electric
flux energy is exponentially suppressed. Where no error bars on the Monte
Carlo data are visible they are smaller than the size of the data points.}
\label{fig:fig4}
\end{figure}

After correcting for lattice artefacts\site{Latt}, the (semi-)analytic results
agree extremely well with the best lattice data\site{Mitt} (with statistical
errors of 2\% to 3\%) up to a volume of about .75 fermi, or about five times
the inverse scalar glueball mass. In figure 4 we present the comparison for a
lattice of spatial size $4^3$. Monte Carlo data\site{Mitt} are most accurate
for this lattice size. For more detailed comparisons see ref.\site{Latt}.
The analytic results below $z=0.95$ are due to L\"uscher and
M\"unster\site{LuMu}, which is where the spectrum is insensitive to the
identifications at the boundary of $\Lm$. 

Most conspicuously the tensor state in finite volumes is split in a doublet 
$E$, with a mass that is roughly 0.9 times the scalar $A_1$ mass and a triplet 
$T_2$ with a mass of roughly 1.7 times the scalar mass. Note that the 
multiplicity weighted average is approximately 1.4 times the scalar mass, 
agreeing well with what was found at large volumes from lattice data\site{Mitt}.

Apart from the corrections for the lattice artefacts, generalisation to SU(3) 
was established by Vohwinkel\site{Voh2}, with qualitatively similar results. 
In large volumes the rotational symmetry should be restored, as is observed 
from lattice simulations. The properties of the fundamental domain restricted 
to the zero-momentum modes for $SU(3)$ can be read off from the results in 
ref.\site{Vba3}. In this reference also the generalisation to arbitrary 
gauge groups is discussed. 

At large volumes extra degrees of freedom start to behave non-perturbatively.
To demonstrate this, the minimal barrier height that separates two vacuum
valleys that are related by gauge transformations with non-trivial winding
number
\beq
\nu(g)={1\over 24\pi^2}\int_M\Tr((g^{-1}dg)^3),
\eeq
was found to be $72.605/Lg^2$, using the lattice approximation and carefully 
taking the continuum limit\site{Gavb}. As long as the states under consideration
have energies below this value, the transitions over this barrier can be 
neglected and the zero-momentum effective Hamiltonian provides an accurate 
description. One can now easily find for which volume the energy of the level 
that determines the glueball mass (defined by the difference with the 
groundstate energy) starts to be of the order of this barrier height. This 
turns out to be the case for $L$ roughly 5 to 6 times the correlation length 
set by the scalar glueball mass. The situation is sketched in figure 5. 

We expect, as will be shown for the three-sphere, that the boundary of 
the fundamental domain along the path in field space across the barrier (which 
corresponds to the instanton path if we parametrise this path by Euclidean 
time $t$), occurs at the saddle point (which we call a finite volume 
sphaleron) in between the two minima.  The degrees of freedom along this 
tunnelling path go outside of the space of zero-momentum gauge fields and if 
the energy of a state flows over the barrier, its wave functional will no 
longer be exponentially suppressed below the barrier and will in particular 
be non-negligible at the boundary of the fundamental domain. Boundary 
identifications in this direction of field space now become dynamically 
important too. The relevant ``Bloch momentum'' is in this case obviously the
$\theta$ parameter, as wave functionals pick up a phase factor $e^{i\theta}$
under a gauge transformation with winding number one. For many of the
intricacies in describing instantons on a torus we refer to 
ref.\site{Ggsv,Buck,Edin}.
On the three-torus we have therefore achieved a self-contained picture
of the low-lying glueball spectrum in intermediate volumes from first
principles with \un{no free parameters}, apart from the overall scale.

\begin{figure}{\tt}
\vspace{8cm}
\includegraphics{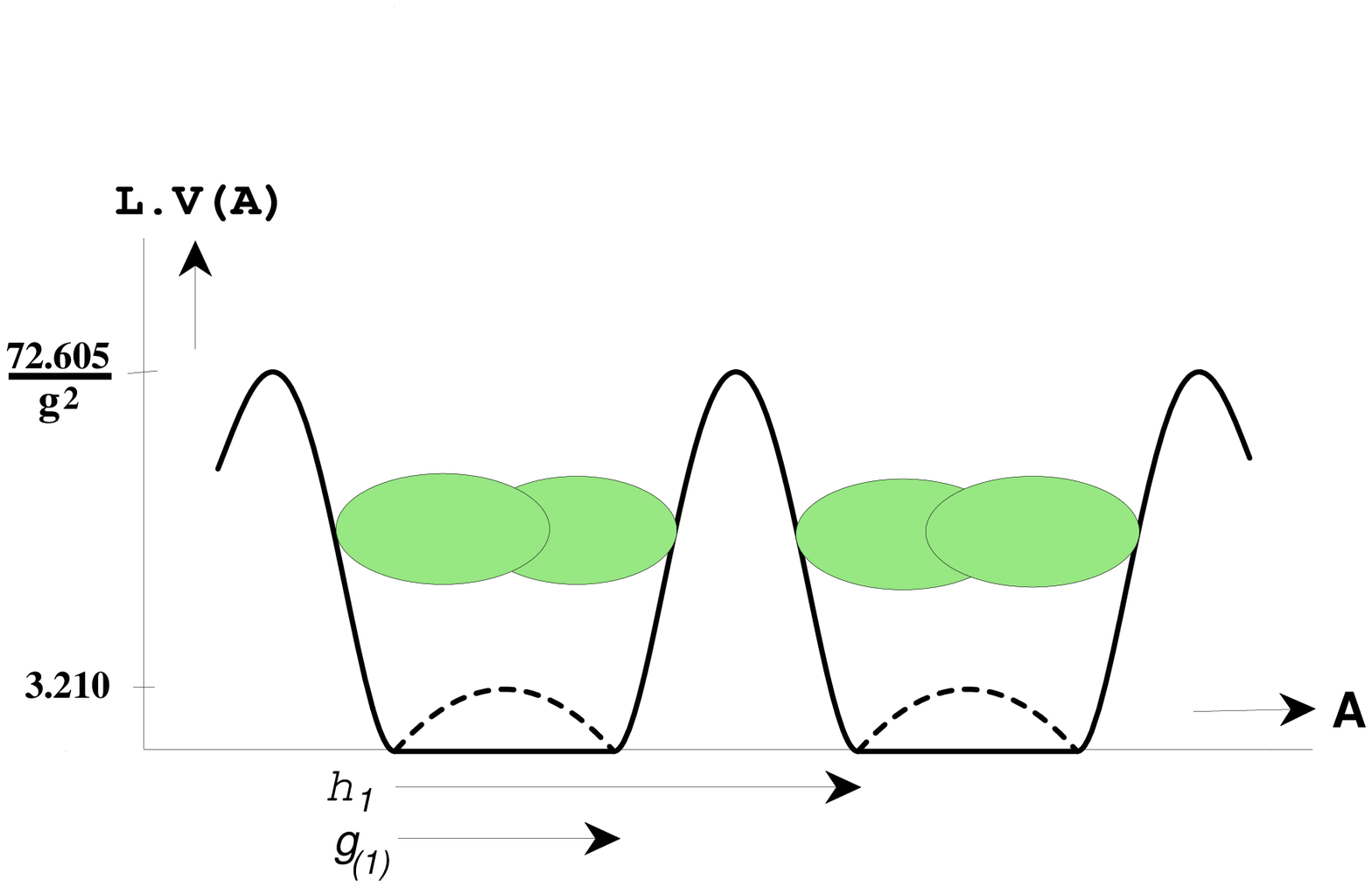}
\caption{``Artistic impression'' of the potential for the three-torus. Shown
are two vacuum valleys, related to each other by a gauge transformation $h_1$
with winding number 1, whose shape in the direc- tion perpendicular to the
valley depends on the position along the valley. The induced one-loop effec-
tive potential, of height $3.210/L$, has degenerate minima related to each 
other by the anti-periodic\break gauge transformations $g_{(k)}$. The classical 
barrier, separating the two valleys, has a height $72.605/Lg^2$.}
\label{fig:fig5}
\end{figure}

\section{GAUGE FIELDS ON THE THREE-SPHERE}

The reason to consider the three-sphere lies in the fact that the conformal
equivalence of $S^3\times\real$ to $\real^4$ allows one to construct instantons
explicitly\site{Vbda,Hoso}. This greatly simplifies the study of how to 
formulate $\theta$ dependence in terms of boundary conditions on the 
fundamental domain, and indeed we will see that for $S^3$ simple enough 
results can be obtained to address this question\site{Vbvd,Vdhe}. The 
disadvantage of the three-sphere is that in large volumes the corrections 
to the glueball masses are no longer exponential\site{Lue1}. 

We will summarise the formalism that was developed in\site{Vbda}. Alternative 
formulations, useful for diagonalising the Faddeev-Popov and fluctuation 
operators, were given in ref.\site{Cutk}. We embed $S^3$ in $\real^4$ by 
considering the unit sphere parametrised by a unit vector $n_\mu$. It is 
particularly useful to introduce the unit quaternions $\sg_\mu$ and their 
conjugates $\bar{\sg}_\mu = \sg^\dagger_\mu$ by
\beq
  \sg_\mu = ( \id ,i \vec{\tau}), \hspace{1.5cm}
  \bar{\sg}_\mu = ( \id,- i \vec{\tau}).
\eeq
They satisfy the multiplication rules
\beq
 \sg_\mu \sgbar_\nu = \eta^\al_{\mu \nu} \sg_\al, \hspace{1.5cm}
  \sgbar_\mu \sg_\nu = \bar{\eta}^\al_{\mu \nu} \sg_\al,
\eeq
where we used the 't Hooft $\eta$ symbols\site{Tho2}, generalised slightly to
include a component symmetric in $\mu$ and $\nu$ for $\al=0$. We can use
$\eta$ and $\bar{\eta}$ to define orthonormal framings\site{Lue2} of $S^3$,
which were motivated by the particularly simple form of the instanton
vector potentials in these framings. The framing for $S^3$ is
obtained from the framing of $\real^4$ by restricting in the following
equation the four-index $\al$ to a three-index $a$
(for $\al = 0$ one obtains the normal on $S^3$):
\beq
  e^\al_\mu = \eta^\al_{\mu \nu} n_\nu , \hspace{1.5cm}
  \bar{e}^\al_\mu = \bar{\eta}^\al_{\mu \nu} n_\nu.
\eeq
Note that $e$ and $\bar{e}$ have opposite orientations. Each framing defines
a differential operator and associated (mutually commuting) angular
momentum operators $\vec{L}_1$ and $\vec{L}_2$:
\beq
  \pr^i = e^i_\mu \frac{\pr}{\pr x^\mu},\quad L_1^i = \frac{i}{2}~\pr^i\quad,
  \quad\bar{\pr}^i = \bar{e}^i_\mu \frac{\pr}{\pr x^\mu},\quad
  L_2^i = \frac{i}{2}~\bar{\pr}^i.
\eeq
It is easily seen that $\Lkw = \vec{L}_2^2$, which has eigenvalues
$l(l+1)$, with $l=0,\half,1,\cdots$.

The (anti-)instantons\site{Bpst} in these framings, obtained from those
on $\real^4$ by interpreting the radius in $\real^4$ as the
exponential of the time $t$ in the geometry $S^3 \times \real$, become
\beq
  A_0 = \frac{\vec{\veps} \cdot \vec{\sg}}{2 ( 1 + \veps \cdot n )}
  , \hspace{1.5cm}
  \vec A = \frac{\vec{\sg} \wedge \vec{\veps} -( u + \veps \cdot n )
  \vec{\sg}} {2 ( 1 + \veps \cdot n )},\label{vecA}
\eeq
where
\beq
   u = \frac{ 2 s^2}{1 + b^2 + s^2} , \hspace{1.5cm}
   \veps _\mu = \frac{2 s b_\mu}{1 + b^2 + s^2} , \hspace{1.5cm}
   s = \lm e^t.
\eeq
Here $\vec\veps$ and $\vec A$ are defined with respect to the
framing $e^a_\mu$ for instantons and with respect to the framing
$\bar{e}^a_\mu$ for anti-instantons.
The instanton describes tunnelling from $A = 0$ at $t = - \infty$ to
$A_a = - \sg_a$ at $t = \infty$, over a potential barrier at $t=0$
that is lowest when $b_\mu \equiv 0$. This configuration corresponds to a
sphaleron\site{Klma}, i.e.\ the vector potential $A_a = -\half\sg_a$
is a saddle point of the energy functional with one unstable mode,
corresponding to the direction ($u$) of tunnelling. At $t = \infty$,
$A_a = - \sg_a$ has zero energy and is a gauge copy of $A_a = 0$ by a
gauge transformation $g = n \cdot \sgbar$ with winding number one.

We will be concentrating our attention to the modes that are degenerate
in energy to lowest order with the modes that describe tunnelling through
the sphaleron and "anti-sphaleron". The latter is a gauge copy by a gauge
transformation $g = n \cdot \sg$ with winding number $-1$ of the sphaleron.
The two dimensional space containing the tunnelling paths through these
sphalerons is consequently parametrised by $u$ and $v$ through
\beq
  A_\mu(u,v)=\left(-u e^a_\mu-v\bar{e}^a_\mu \right)\frac{\sg_a}{2}.
\eeq
The gauge transformation with winding number $-1$ is easily seen to
map $(u,v) = (w,0)$ into $(u,v) = (0,2-w)$.
The 18 dimensional space is defined by
\beq
  A_\mu(c,d)=\left(c^a_i  e^i_\mu+d^a_j\bar{e}^j_\mu \right)
  \frac{\sg_a}{2}=A_i(c,d)e_\mu^i.
  \label{Acddef}
\eeq
The $c$ and $d$ modes are mutually orthogonal and satisfy
the Coulomb gauge condition:
\beq
  \pr_i A_i(c,d) = 0.
\eeq
This space contains the $(u,v)$ plane through $c^a_i = -u \dl^a_i$ and
$d^a_i = -v \dl^a_i$. The significance of this 18 dimensional space is that the
energy functional\site{Vbda}
\beq
  \Ss{V}(c,d) \equiv - \int_{S^3} \frac{1}{2} \tr(F_{ij}^2)
  = \Ss{V}(c) + \Ss{V}(d) + \frac{2 \pi^2}{3}
   \left\{ (c^a_i)^2 (d^b_j)^2 - (c^a_i d^a_j)^2 \right\}\label{pot}
,\eeq
\beq
  \Ss{V}(c) = 2 \pi^2 \left\{ 2 (c^a_i)^2 + 6 \det c +
  \frac{1}{4}[(c^a_i c^a_i)^2 - (c^a_i c^a_j)^2 ] \right\} ,
\eeq
is degenerate to second order in $c$ and $d$. Indeed, the quadratic
fluctuation operator \Ss{M} in the Coulomb gauge, defined by
\bea
  - \int_{S^3} \frac{1}{2} \tr(F_{ij}^2)
  &=& \int_{S^3} \tr(A_i \Ss{M}_{ij} A_j) + \Order{A^3},\nonumber \\
  \Ss{M}_{ij} &=& 2\vec{L}_1^2\dl_{ij} + 2 \left( \vec{L}_1 + \vec{S}
  \right)^2_{ij},\quad S^a_{ij}=-i \veps_{aij},\label{fluct}
\eea
has $A(c,d)$ as its eigenspace for the (lowest) eigenvalue $4$. These
modes are consequently the equivalent of the zero-momentum modes on the
torus, with the difference that their zero-point frequency does not vanish.

$FP_f(A)$ in eq.~(\ref{FPhalfdef}) is defined as a hermitian
operator acting on the vector space $\Ss{L}$ of functions $g$ over $S^3$
with values in the space of the quaternions $\quat=\{q_\mu\sg_\mu|
q_\mu\in\real\}$. The gauge group $\Ss{G}$ is contained in $\Ss{L}$
by restricting to the unit quaternions:
$\Ss{G}=\{g\in\Ss{L}|g=g_\mu\sg_\mu,g_\mu\in\real,g_\mu g_\mu = 1 \}$.
For arbitrary gauge groups $\Ss{L}$ is defined as the algebra generated
by the identity and the (anti-hermitian) generators of the algebra.
When minimising the same functional over the larger space $\Ss{L}$
one obviously should find a smaller space $\tilde{\Lm} \subset \Lm$.
Since $\Ss{L}$ is a linear space $\tilde\Lm$ can also be specified by
the condition that $FP_f(A)$ be positive,
\beq
  \tilde{\Lm} =
  \{ A \in \Gm | \ip{g, FP_f(A)~g}\geq0,\ \forall g\in\Ss{L} \}.
  \label{Lmtildef}
\eeq
As for the Gribov horizon, the boundary of $\tilde\Lm$ is therefore determined
by the location where the lowest eigenvalue vanishes. For the $(c,d)$ space
it can be shown\site{Vbvd} that the boundary $\partial\tilde{\Lm}$ will touch
the Gribov horizon $\partial\Om$. This establishes the
existence of singular points on the boundary of the fundamental domain due
to the inclusion $\tilde{\Lm} \subset \Lm \subset \Om$.
By showing that the fourth order term in eq.~(\ref{Xexpansie}) is positive
(see app.~A of ref.\site{Vbvd}) this is seen to correspond to the situation
as sketched in figure 2.

To simplify the notation, write $FP_f=FP_\half$ and $FP=FP_1$, with the indices
related to the isospin. The associated generators are 
\beq
\vec T_\half=\half\vec \tau\qquad \mbox{and}\qquad \vec T_1=\half\ad\vec\tau.
\eeq
One can now make convenient use of the 
${\rm SU}(2)^3$ symmetry generated by $\vec L_1$, $\vec L_2$ and $\vec T_t$ to 
calculate explicitly the spectrum of $FP_t(A)$. One has
\beq
  FP_t(A(c,d)) = 4 \Lkw - \frac{2}{t} c^a_i T_t^a L_1^i
   - \frac{2}{t} d^a_i T_t^a L_2^i,
\eeq
which commutes with $\Lkw = \vec{L}_2^2$, but for arbitrary $(c,d)$
there are in general no other commuting operators (except for
a charge conjugation symmetry when $t=\half$). Restricting to
the $(u,v)$ plane one easily finds that
\beq
  FP_t(A(u,v)) = 4 \Lkw + \frac{2}{t} u \vec{L}_1 \cdot \vec{T}_t
    + \frac{2}{t} v \vec{L}_2 \cdot \vec{T}_t,
    \label{FPuvdef}
\eeq
which also commutes with the total angular momentum $\vec{J}_t=\vec{L}_1
+\vec{L}_2+\vec{T}_t$ and is easily diagonalized. Figure 6 summarises
the results for this $(u,v)$ plane and also shows the equal-potential
lines as well as exhibiting the multiple vacua and the sphalerons. As it
is easily seen that the two sphalerons are gauge copies (by a unit winding
number gauge transformation) with equal norm, they lie on $\partial\Lm$,
which can be extend by perturbing around these sphalerons\site{Vbcu}.

\begin{figure*}{\tt}
\vspace{9.5cm}
\includegraphics{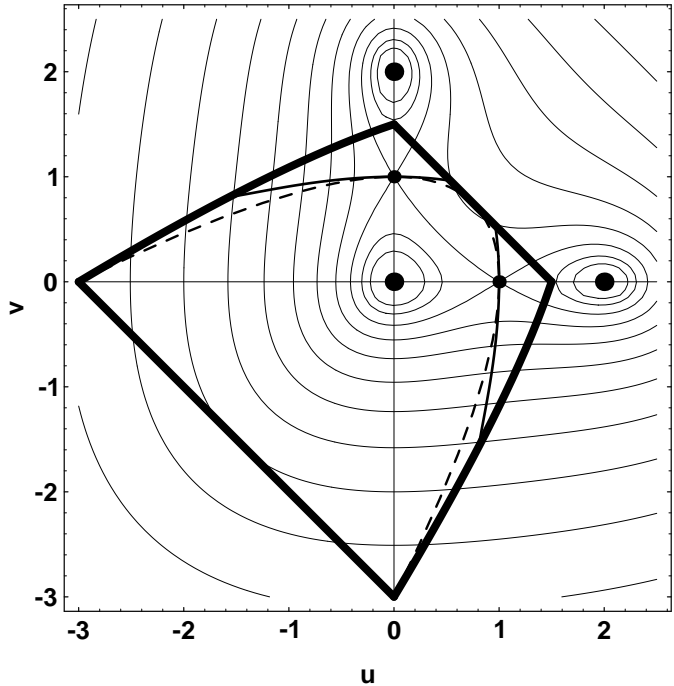}
\caption{Location of the classical vacua (large dots), sphalerons (smaller
dots), the Gribov horizon (fat sections), the boundary of $\tilde\Lm$
(dashed curves) and part of the boundary of the fundamental domain (full
curves). Also indicated are the lines of equal potential in units of
$2^n$ times the sphaleron energy.}
\label{fig:fig6}
\end{figure*}

To obtain the result for general $(c,d)$ one can use the
invariance under rotations generated by $\vec{L}_1$ and $\vec{L}_2$
and under constant gauge transformations generated by $\vec{T}_t$,
to bring $c$ and $d$ to a standard form, or express
$\det\left(FP_t(A(c,d))|_{l = \half}\right)$, which determines the
locations of $\partial\Om$ and $\partial\tilde\Lm$, in terms of invariants.
We define the matrices $X$ and $Y$ by $X^a_b = (c c^t)^a_b$ and
$Y^a_b = (d d^t)^a_b$, which allows us to find
\bea
\det\left(FP_\half (A(c,d))|_{l = \half}\right)&=&
[81 - 18 \Tr (X +Y) + 24 ( \det c + \det d)\nonumber\\
 & & - (\Tr(X - Y))^2 + 2 \Tr((X-Y)^2)]^2.
  \label{halfcd}
\eea
The two-fold multiplicity is due to charge conjugation symmetry.
The expression for $t=1$, that determines the location of the
Gribov horizon in the $(c,d)$ space, is given in app.~B of ref.\site{Vbvd}.
If we restrict to $d=0$ the result simplifies considerably. In that case
one can bring $c$ to a diagonal form $c_i^a=x_i\delta_i^a$. The rotational
and gauge symmetry reduce to permutations of the $x_i$ and simultaneous
changes of the sign of two of the $x_i$. One easily finds the invariant
expression ($\Tr(X)=\sum_i x_i^2$ and $\det c=\prod_i x_i$)
\beq
  \det\left(FP_1(A(c,0))|_{l = \half}\right)
= \left( 2 \det c - 3 \Tr(X) + 27 \right)^4.
\label{deteend0}
\eeq

In figure 7 we present the results for $\Lm$ and $\Om$. In this particular
case, where $d=0$, $\Lambda$ coincides with $\tilde\Lambda$, a consequence of
the convexity and the fact that both the sphalerons (indicated by the dots)
and the edges of the tetrahedron lie on $\partial\Lambda$, the latter also
lying on $\partial\Om$. It is essential that the sphalerons do not lie on
the Gribov horizon and that the potential energy near $\partial\Om$ is
relatively high. This is why we can take the boundary identifications near
the sphalerons into account without having to worry about singular boundary
points, as long as the energies of the low-lying states will be not much
higher than the energy of the sphaleron. It allows one to study the
glueball spectrum as a function of the CP violating angle $\theta$, but
more importantly it incorporates for $\theta=0$ the noticeable influence
of the barrier crossings, i.e. of the instantons. 

\begin{figure*}{\tt}
\vspace{7cm}
\includegraphics{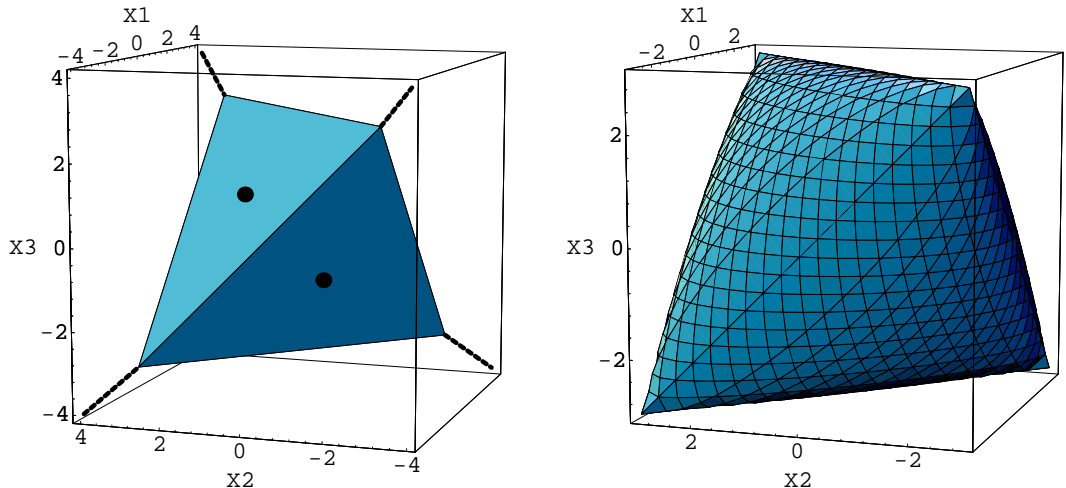}
\caption{The fundamental domain (left) for constant gauge fields on $S^3$, with
respect to the ``instanton'' framing $e_\mu^a$, in the diagonal representation
$A_a=x_a\sg_a$ (no sum over $a$). By the dots on the faces we indicate the
sphalerons, whereas the dashed lines represent the symmetry axes of the
tetrahedron. To the right we display the Gribov horizon, which encloses
the fundamental domain, coinciding with it at the singular boundary points
along the edges of the tetrahedron.}
\label{fig:fig7}
\end{figure*}

\atable{Numerical coefficients for $\Ss{V}^{(1)}_{\rm efff}$}{
\begin{tabular}{l@{ = }r@{.}l|l@{ = }r@{.}l} \hline
$\kappa_1$ & $-0$&$2453459985179565 $ &
$\kappa_2$ & $ 3$&$66869179814223   $\\
$\kappa_3$ & $ 0$&$500703203096610  $ &
$\kappa_4$ & $-0$&$839359633413003  $\\
$\kappa_5$ & $-0$&$849965412245339  $ &
$\kappa_6$ & $-0$&$06550330854836428$\\
$\kappa_7$ & $-0$&$3617122159967145 $ &
$\kappa_8$ & $-2$&$295356861354712  $\\ \hline
\end{tabular}}

An effective Hamiltonian for the $c$ and $d$ modes is derived from 
the one-loop effective action\site{Vdhe}. To lowest order it is 
given by 
\beq
H=-\frac{g^2(R)}{4\pi^2R}\left(\left(\frac{\partial}{\partial c_i^a}\right)^2+
\left(\frac{\partial}{\partial d_i^a}\right)^2\right)
+\frac{1}{g^2(R)R}\Ss{V}(c,d)+\frac{1}{R}\Ss{V}^{(1)}_{\rm eff}(c,d),
\eeq
where $g(R)$ is the running coupling constant (related to the MS running
coupling by a finite renormalisation, such that kinetic term above has no
corrections). The one-loop correction to the effective potential\site{Vdhe}
is given by (for $\kappa_i$ see table 1):
\bea
  \Ss{V}^{(1)}_{\rm eff}(c,d)&=&\Ss{V}^{(1)}_{\rm eff}(c)+
  \Ss{V}^{(1)}_{\rm eff}(d)+\kappa_7(c^a_i)^2 (d^b_j)^2+
  \kappa_8(c^a_i d^a_j)^2,\\ 
  \Ss{V}^{(1)}_{\rm eff}(c)&=&\kappa_1(c^a_i)^2+\kappa_2\det c+
  \kappa_3(c^a_i c^a_i)^2+\kappa_4(c^a_i c^a_j)^2+\kappa_5(c^a_i)^2\det c+
  \kappa_6(c^a_i c^a_i)^3.\nonumber
\label{potonel}
\eea
Errors due to an adiabatic approximation
are not necessarily suppressed by powers of the coupling constant.
Nevertheless, one expects to achieve an approximate understanding of
the non-perturbative dynamics in this way\site{Vdhe}.

The boundary conditions are chosen so as to coincide with the appropriate 
boundary conditions near the sphalerons, but such that the gauge and (left 
and right) rotational invariances are not destroyed. Projections on the 
irreducible representations of these symmetries turned out to be essential to 
reduce the size of the matrices to be diagonalised in a Rayleigh-Ritz analysis.
Remarkably all this could be implemented in a tractable way\site{Vdhe}. Results
are summarised in figure 8. One of the most important features is that the 
$0^-$ glueball is (slightly) lighter than the $0^+$ in perturbation theory, 
but when including the effects of the boundary of the fundamental domain, 
setting in at $f\equiv g^2/2\pi^2\sim 0.2$, the $0^-/0^+$ mass ratio rapidly 
increases. Beyond 
$f\sim 0.28$ it can be shown that the wave functionals start to feel parts of 
the boundary of the fundamental domain which the present calculation is not 
representing properly\site{Vdhe}. This value of $f$ corresponds to a 
circumference of roughly 1.3 fm, when setting the scale as for the torus, 
assuming the scalar glueball mass in both geometries at this intermediate 
volume to coincide. 

\begin{figure}{\tt}
\vspace{7.3cm}
\includegraphics{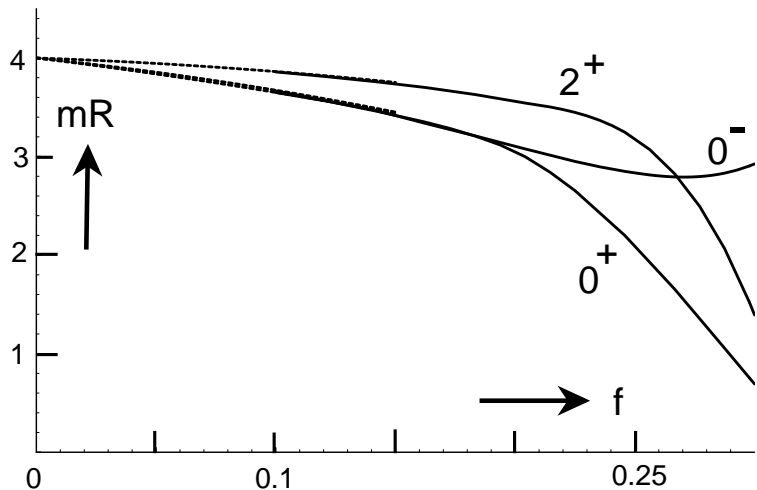}
\caption{The full one-loop results for the masses of scalar, tensor and odd
glueballs on $S^3$ as a function of $f=g^2(R)/2\pi^2$ for $\theta=0$. The
dashed lines correspond to the perturbative result.}
\label{fig:fig3a}
\end{figure}

\section{DISCUSSION}

We have analysed in detail the boundary of the fundamental domain for SU(2) 
gauge theories on the three-torus and three-sphere. It is important to note 
that it is necessary to divide $\Ss{A}$ by the set of \un{all} gauge 
transformations, including those that are homotopically non-trivial, to get 
the physical configuration space. All the non-trivial topology is then 
retrieved by the identifications of points on the boundary of the fundamental 
domain.

As we already mentioned in the introduction, the knowledge of the boundary 
identifications is important in the case that the wave functionals spread 
out in configuration space to such an extent that they become sensitive to 
these identifications. This happens at large volumes, whereas at very small 
volumes the wave functional is localised around $A=0$ and one need not worry 
about these non-perturbative effects. That these effects can be dramatic, 
even at relatively small volumes (above a tenth of a fermi across), was 
demonstrated for the case of the torus. However, for that case the structure 
of the fundamental domain (restricted to the abelian zero-energy modes) is a 
hypercube and deviates considerably from the fundamental domain of the 
three-sphere. Nevertheless, the spectrum for the sphere is compatible with 
that for the torus in volumes around one fermi across\site{Mite}, with 
$m(2^+)/m(0^+)\sim 1.5$ and $m(0^-)/m(0^+)\sim 1.7$. 

It should be noted that the shape of $\Lm$ is independent of $L$ if the gauge
field is expressed in units of $1/L$. Suppose that the coupling constant will 
grow without bound. This would make the potential irrelevant and makes the 
wave functional spread out over the whole of field space (which could be seen 
as a strong coupling expansion). If the kinetic term would have been trivial 
the wave functionals would be ``plane waves'' on a space with complicated 
boundary conditions. In that case it seems unavoidable that the infinite volume
limit would depend on the geometry (like $T^3$ or $S^3$) that is scaled-up to 
infinity. With the non-triviality of the kinetic term this conclusion cannot 
be readily made and our present understanding only allows comparison in 
volumes around one cubic fermi. However, one way to avoid this undesirable 
dependence on the geometry is that the vacuum is unstable against domain 
formation. As periodic subdivisions are space filling on a torus, this seems 
to be the preferred geometry to study domain formation. In a naive way it will 
give the correct string tension (flux conservation tells us to ``string'' the
domains that carry electric flux) and tensor to scalar mass ratio (averaging 
over the orientations of the domains is expected to lead to a multiplicity 
weighted average of the $T_2$ and $E$ masses). Furthermore, the natural 
dislocations of such a domain picture are gauge dislocations. The point-like 
gauge dislocations in four dimensions are instantons and in three dimensions 
they are monopoles. Their density is expected to be given roughly as one per 
domain (with a volume of around 0.5 cubic fermi). Also the coupling constant 
will stop running at the scale of the domain size. We have discussed this 
elsewhere and refer the reader to refs.\site{Vba3,Edin,Vbk2} for further 
details. 

{\referencestyle

}
\end{document}